\documentstyle{article}
\title{\vspace*{-2.1cm}\hspace*{8.7cm} {\large gr-qc/9311020,
CGPG-93/11-2\vspace*{.9cm}}\\
\Large{{\bf Reality Conditions for Lorentzian and Euclidean Gravity in the
Ashtekar Formulation}}}
\author{\\Guillermo A. Mena Marug\'an\vspace*{.6cm}\\
Center for Gravitational Physics and Geometry, Pennsylvania State
University,\\ 104 Davey Laboratory, University Park, PA 16802-6300, USA.
\vspace*{.4cm}\\ On leave from: {\it Instituto de Matem\'aticas y F\'{\i}sica
Fundamental},\\ {\it C.S.I.C., Serrano 121, 28006 Madrid, Spain.}\\ }
\date{November, 1993}

\begin{document}

\maketitle
\large
\setlength{\baselineskip}{.825cm}

\begin{center}
{\bf Abstract}
\end{center}

Using Ashtekar variables, we analyze Lorentzian and Euclidean gravity in
vacuum up to a constant conformal transformation. We prove that the
reality conditions are invariant under a Wick rotation of the time, and
show that the compatibility of the algebra of commutators and constraints
with the involution defined by the reality conditions restricts the
possible values of the conformal factor to be either real or purely
imaginary. In the first case, one recovers real Lorentzian general
relativity. For purely imaginary conformal factors, the classical theory
can be interpreted as real Euclidean gravity. The reality conditions
associated with this Euclidean theory demand the hermiticity of the
Ashtekar connection, but the densitized triad is represented by an
anti-Hermitian operator. We also demonstrate that the Euclidean and
Lorentzian sets of reality conditions lead to inequivalent quantizations
of full general relativity. As a consequence, it seems impossible to
obtain Lorentzian physical predictions from the quantum theory constructed
with the Euclidean reality conditions.

\vspace*{.24cm}

\newpage

\section {Introduction}

The Ashtekar formulation of general relativity [1-4] provides one of the
most promising approaches to construct a consistent theory of quantum
gravity. The Ashtekar gravitational variables, a densitized triad and a
canonically conjugate complex connection [1-3], appear to be specially
well-suited to deal with the type of problems that one encounters in
quantizing the gravitational field. In particular, the use of connections
leads in a natural way to the loop representation for quantum gravity
[5-7], in which much progress has been obtained during the last five years
[7,8].

In the search of a quantum theory of gravity, the introduction of the
Ashtekar variables has been complemented with a systematic quantization
program [4] that, in addition to the non-perturbative canonical
quantization scheme proposed by Dirac [9], includes the mathematical
machinery needed to determine the inner product in the space of physical
states [3,10]. One first selects an overcomplete set of complex functions
on phase space that is closed under Poisson brackets. For pure gravity,
for instance, the Ashtekar variables provide a set with these properties.
The selected set is promoted to an abstract $\ast$-algebra of elementary
operators in such a way that the Poisson brackets are straightforwardly
translated into commutators (up to leading order in $\hbar$) and the
complex conjugation relations between classical variables are captured in
the involution operation. The corresponding $\ast$-relations between
elementary operators are usually called reality conditions [3,11]. The
abstract algebra of basic operators is represented then on a chosen vector
space, with the physical states annihilated by all the first-class
constraints of the theory [3,9]. At this point, one should find a
sufficiently large number of observables for the system (i.e., operators
that commute with all the constraints), derive the $\ast$-relations
between observables that are induced by the reality conditions, and
promote these relations to adjointness requirements in the Hilbert space
of physical states [4]. Mathematically, these adjointness conditions
determine the inner product in the quantum theory if the number of
observables known is large enough [10]. Physically, these requirements
guarantee that the spectrum (and then every quantum measurement) of any
Hermitian observable is real;  where we understand that an operator is
Hermitian when it coincides with its $\ast$-conjugate.

Notice that the reality conditions between operators that are not
observables cannot be promoted to adjointness conditions, because the
action of these operators is not well-defined in the space of physical
states. If one were able to isolate the physical degrees of freedom of the
theory, the reality conditions on the non-physical degrees could be either
imposed as second-class constraints before quantization or obviated by
simply quantizing the associated reduced phase space. Since all complex
functions on the reduced phase space are observables, the reality
conditions for the reduced theory can always be promoted to adjointness
requirements. The approach adopted in the non-perturbative quantization
program is nevertheless more general, in the sense that one does not
assume that the reduced phase space is known to carry out the
quantization. However, one must keep in mind that, at the end of the day,
only reality conditions on observables will play a decisive role in
determining the quantum theory.

Among the problems that one finds in implementing the non-perturbative
canonical quantization program, the imposition of the reality conditions
that correspond to Lorentzian gravity in the Ashtekar formulation appears
as one of the main technical difficulties that has to be solved in order
to complete the quantization of general relativity. We recall that the
Ashtekar connection is genuinely complex, its real part being given by the
connection that is compatible with the triad, and its imaginary part by
the extrinsic curvature [2,3]. As a consequence, even if one could
determine the whole space of physical states and a complete set of
observables for gravity, the adjointness requirements derived from the
reality conditions on the complex Ashtekar connection and the densitized
triad might be very difficult to compute and implement. The problems with
reality conditions are also transmitted to the loop representation, where
the $\ast$-relations for the elementary operators [6,7], constructed using
Ashtekar variables, are not even explicitly known.

Opposite to the situation in Lorentzian gravity, it has been pointed out
that the reality conditions for the Ashtekar variables should be very
simple in Euclidean general relativity [3]. The basic remark is that,
under a Wick rotation of the time [12,13], the Lorentzian extrinsic
curvature transforms into $i$ times its Euclidean counterpart, the latter
being real for Euclidean spaces. Therefore, the Ashtekar connection turns
out to be real for Euclidean gravity, and one expects it to be represented
by a Hermitian operator. The reality conditions for the densitized triad,
on the other hand, should guarantee that the classical 3-metric is real.
Given the apparent simplicity of these requirements, it seems natural to
ask which are exactly the reality conditions for Euclidean gravity in the
Ashtekar formulation, whether it is possible to adopt such conditions to
quantize general relativity and if one can extract Lorentzian physical
predictions from the quantum theory so obtained. The aim of this paper is
to investigate the answers to these questions, and clarify the sense in
which one can refer to Lorentzian and Euclidean gravity as two quantum
theories built out of different sets of reality conditions.

In Sec. II, we argue that the reality conditions for the Euclideanized
theory of general relativity obtained by a Wick rotation of the time
coincide with those corresponding to Lorentzian gravity, so that both
classical theories lead in fact to the same quantization. In Sec. III, we
analyze a family of classical theories that describe either Lorentzian or
Euclidean gravity up to a constant conformal transformation. We prove
that, if one requires that the real densitized triad is Hermitian, the
consistency of the $\ast$-operation with the algebra of elementary
operators and constraints restricts the complex conformal factor to be
either real or purely imaginary. The classical theories associated with
these two different values of the conformal factor can be interpreted as
real Lorentzian and real Euclidean general relativity, respectively. Sec.
IV deals with the Euclidean set of reality conditions. We show there that,
under very general assumptions, the quantum theories selected by the
Lorentzian and Euclidean sets of reality conditions result in being
inequivalent. Finally, we summarize the results in Sec.V.

\section{Lorentzian Gravity and Wick Rotation}

The Ashtekar gravitational variables can be taken as a densitized triad,
$\tilde{E}^a_{\;i}$, and a SO(3) connection, $A_a^{\;i}$, both of them
defined over a 3-manifold $\Sigma$ [3]. In the following, the spatial and
SO(3) indices will be denoted by lower case Latin letters from the
beginning and the middle of the alphabet, respectively. The Ashtekar
connection is canonically conjugate to the densitized triad:
\begin{equation} \{A_a^{\;i}(x),\tilde{E}^b_{\;j}(y)\}=i\delta^b_a
\delta^i_j \delta^3(x,y),\end{equation}
the rest of Poisson brackets between the Ashtekar variables being equal to
zero. In the sector of non-degenerate metrics, the Ashtekar variables can
be written in terms of the extrinsic curvature, $K_{ab}$, and the triad,
$e^a_{\;i}$ [1-3],
\begin{equation}\tilde{E}^a_{\;i}=e^a_{\;i}\;q(e),\;\;\;A_a^{\;i}=
\Gamma_a^{\;i}(e)-i\,K_{ab}e^{bi},\end{equation}
where the SO(3) indices are raised and lowered with the metric
$\eta^{ij}=(1,1,1)$, \linebreak $q^{ab}=e^a_{\;i}e^{bj}$ is the inverse
3-metric, $q(e)=({\rm det}q_{ab})^{1/2}$, and $\Gamma_a^{\;i}(e)$ is the
SO(3) connection compatible with the triad:
\begin{equation}\Gamma_a^{\;i}=\frac{1}{2}\epsilon^{ijk}
E_{_{_{\!\!\!\!\!\!\sim}}\;jb}\left(-\partial_a
\tilde{E}^b_{\;k}+\Gamma^b_{\;\;ca}\tilde{E}^c_{\;k}\right).\end{equation}
Here, $\epsilon^{ijk}$ is the anti-symmetric symbol,
$E_{_{_{\!\!\!\!\!\!\sim}}\;ia}$ denotes the inverse of the densitized
triad, and $\Gamma^a_{\;\;bc}$ are the Christoffel symbols [14]:
\begin{equation} \Gamma^a_{\;\;bc}=\frac{1}{2}q^{ad}\left(\partial_cq_{db}+
\partial_bq_{dc}-\partial_dq_{bc}\right).\end{equation}
Finally, the extrinsic curvature can be expressed in terms of the lapse
function, $N$, the shift functions, $N^a$, and the time derivatives of the
3-metric [14]
\begin{equation} K_{ab}=\frac{1}{2N}\left(\partial_tq_{ab}-N_{(a;b)}\right),
\end{equation}
with $N_{(a;b)}$ the symmetrized covariant derivative (determined by the
triad) of\linebreak $N_a=q_{ab}N^b$.

In the Ashtekar formulation, the constraints for pure gravity adopt the simple
expressions
\begin{equation} {\cal G}_i\equiv {\cal D}_a \tilde{E}^a_{\;i}=\partial_a
\tilde{E}^a_{\;i}+\epsilon_{ij}^{\;\;\;\,k}A_a^{\;j}\tilde{E}^a_{\;k}=0,
\end{equation}
\begin{equation} {\cal V}_a\equiv \tilde{E}^b_{\;i}F_{ba}^{\;\;\;\;i}=0,
\end{equation}
\begin{equation} {\cal S}\equiv \epsilon^{ij}_{\;\;\;\,k}\tilde{E}^a_{\;i}
\tilde{E}^b_{\;j}F_{ab}^{\;\;\;\;k}=0.\end{equation}
$F_{ab}^{\;\;\;\;i}$ is the curvature of the SO(3) connection $A_a^{\;i}$:
\begin{equation} F_{ab}^{\;\;\;\;i}=\partial_a A_b^{\;i}-\partial_b A_a^{\;i}+
\epsilon^i_{\;\,jk}A_a^{\;j}A_b^{\;k}.\end{equation}
Constraints (6-8) are usually referred to as the Gauss law, the vector
constraint and the scalar constraint, respectively.

It is well known that, for Lorentzian gravity, the reality conditions on the
Ashtekar elementary operators can be written as [3]
\begin{equation} (\hat{\tilde{E}\;}\!\,^a_{\;i})^{\ast}=
\hat{\tilde{E}\;}\!\,^a_{\;i},\;\;\;\;(\hat{A}_a^{\;i})^{\ast}=
-\hat{A}_a^{\;i}+2\Gamma_a^{\;i}(\hat{\tilde{E}\;}).\end{equation}
These reality conditions are the straightforward translation to the
algebra of operators of the requirements that the classical densitized
triad must be real (so that the metric
$\tilde{\tilde{q}}^{ab}=\tilde{E}^a_{\;i}\tilde{E}^{bi}$ is positive) and
the real part of the classical Ashtekar connection be given by the SO(3)
connection compatible with the triad, which implies in turn that the
extrinsic curvature $K_{ab}$ in Eq. (2) has to be real. The second of
reality conditions (10) is highly non-polynomial in the densitized triad.
Although it is possible to recast these reality conditions in a polynomial
form [11], we will use in the following expressions (10) as the Lorentzian
reality conditions to simplify our calculations.  On the other hand, if
one insists that the connection $\Gamma_a^{\;i}$ must be well-defined in
terms of the densitized triad, one can always restrict his attention to
the sector of non-degenerate metrics in a consistent way.

The fact that the reality conditions for Lorentzian gravity can be
obtained as a direct translation of the complex conjugation relations
between classical variables may lead us to misunderstand what reality
conditions should be in more general cases.  We recall that reality
conditions are simply given by an involution in the algebra of elementary
operators. When promoted to adjointness requirements, these conditions
guarantee that the spectra of certain operators are real, and it is in
this sense that they are related to complex conjugation conditions. This
implies by no means that the $\ast$-relations in the algebra of operators
should be exactly the operator version of the classical complex
conjugation relations.

To explore this subject in more detail, let us consider the Euclideanized
version of Lorentzian general relativity obtained by a Wick rotation of
the time, rotation that can be accomplished by the substitution
$N\rightarrow -iN$, with $N$ a real lapse function. This Euclidean theory
is usually called Euclidean gravity in the literature [12,13,15]. However,
we will keep the terminology ``Euclideanized Lorentzian gravity'' to refer
to it; the reasons to adopt this name will become obvious in a moment.
Extrapolating the results of Euclidean field theory [16], it has been
frequently assumed that the quantum theory of Lorentzian general
relativity can be reconstructed from the quantization of its Euclideanized
version [12,13]. Therefore, one would expect those two quantum theories to
be somehow equivalent. Actually, under a Wick rotation of the time
($N\rightarrow -iN$), the basic Poisson brackets (1) and the gravitational
constraints (6-8) remain invariant. The only important change is that the
Lorentzian extrinsic curvature transforms into $i$ times its Euclidean
counterpart, that is real in the Euclidean regime:
\begin{equation} (K_E)_{ab}=\frac{1}{2N}\left(\partial_{\tau}q_{ab}-N_{(a;b)}
\right),\end{equation}
where $\tau$ is the Euclidean time. Thus, the Euclidean Ashtekar connection
takes the manifestly real expression
\begin{equation} (A_E)_a^{\;i}=\Gamma_a^{\;i}(e)+(K_E)_{ab}
e^{bi}.\end{equation}

At this point, one is tempted to assert that the reality conditions for
Euclideanized Lorentzian gravity are simply that the operators
$\hat{\tilde{E}\;}\!\,^a_{\;i}$ and $\hat{A}_a^{\;i}$ must be Hermitian.
However, these conditions are inconsistent with the algebra of commutators
derived from Eq. (1),
\begin{equation} [ \hat{A}_a^{\;i}(x),\hat{\tilde{E}\;}\!\,^b_{\;j}(y)]=-\hbar
\delta^b_a\delta^i_j\delta^3(x,y),\end{equation}
and the general properties of any involution
\begin{equation} \left(\lambda \hat{X}\right)\!^{\ast}=\bar{\lambda}
\hat{X}^{\ast},\end{equation}
\begin{equation} \left(\hat{X}\hat{Y}\right)\!^{\ast}=\hat{Y}^{\ast}
\hat{X}^{\ast},\end{equation}
where $\lambda$ is a complex number, $\hat{X}$ and $\hat{Y}$ are two generic
operators, and $^{^{-\!-}}$ denotes complex conjugation. Note that, in
particular, Eq. (15)  implies that $\hat{1}^{\ast}=\hat{1}$.

To prove such an inconsistency, it is enough to take the $\ast$-conjugate
of the commutator (13) and use Eqs. (14,15). If both the densitized triad
and the Ashtekar connection were Hermitian, one would arrive at the
conclusion that their commutator should be given by the right-hand side of
Eq. (13) but with a flip of sign, in clear contradiction with our original
assumptions.

In fact, it is a standard result of Euclidean field theory that the
$\ast$-relations for the Euclidean fields are given by the composition of
the time reversal and the complex conjugation operations [16]. Choosing
the manifold $\Sigma$ as the zero time section of the four-dimensional
space, and taking into account that the densitized triad remains invariant
under time reversal, while the Euclidean extrinsic curvature changes its
sign, we conclude that (if the results of Euclidean field theory are
applicable to general relativity) the reality conditions for Euclideanized
Lorentzian gravity coincide with the Lorentzian conditions (10).
Therefore, the $\ast$-algebra of elementary operators and constraints for
Lorentzian general relativity and its Euclideanized version are exactly
the same, and both classical theories lead indeed to the same
quantization.

It may seem strange that a quantum theory can describe simultaneously two
classical theories whose associated equations of motion are of Lorentzian
and Euclidean type, respectively. We notice, nevertheless, that both kinds
of dynamics are related by an analytic continuation of the time
coordinate. From the point of view of the algebraic structures and
constraints that determine the quantum theory there is however no concept
of time a priori. It is only after introducing a time parameter in the
quantum theory (an intrinsic time) that one can recover the notion of
dynamical evolution and, presumably, the classical equations of motion in
a certain limit. As it usually happens in the WKB approximation employed
in quantum cosmology [17-19], we expect then that, for real values of the
introduced time parameter, one can regain Lorentzian dynamics, while, for
imaginary times obtained by means of an analytic continuation, the
dynamical regime should become Euclidean.

\section{Conformally Lorentzian and Euclidean Gravity}

We are interested in determining whether it is possible to find a set of
reality conditions, other than those given by Eq. (10), that correspond in
some sense to a classical Euclidean theory of gravity and such that, in
particular, the Ashtekar connection is represented by a Hermitian
operator. To explore this topic, we will analyze in this section a family
of classical theories that, at least in vacuum, lead to either Lorentzian
or Euclidean dynamics.

We will consider those sections of the complex phase space of general
relativity for which the 4-metric is either Lorentzian or Euclidean,
modulo a constant conformal transformation:
\begin{equation} ds^2=\Omega^2 (ds_P)^2.\end{equation}
Here, $\Omega$ is a complex constant whose absolute value is equal to the
unity (otherwise, $|\Omega|$ can be absorbed into the
line element $(ds_P)^2$),
\begin{equation} \Omega=e^{i\Theta},\;\;\;\;\Theta\in[0,2\pi).\end{equation}
The 3-metric $(q_P)_{ab}$ determined by $(ds_P)^2$ will be assumed to be
real and non-negative. The lapse and shift functions associated with
$(ds_P)^2$ will be denoted, respectively, by $N_P$ and $N_P^{\;a}$.
Although one can take $N_P$ and $N_P^{\;a}$ to be complex in general, we
will restrict our attention to the cases $N_P^{\;a},N_P^{\;}\in
I\!\!\!\,R$, corresponding to Lorentzian metrics, and $N_P^{\;a}$ real and
$N_P$ imaginary, for which the line element $(ds_P)^2$ results in being
Euclidean.

We notice that the constant conformal factor $\Omega^2$ will decouple from
the equations of motion in pure gravity. As a consequence, the evolution
of the real metric $(q_P)_{ab}$ will be completely consistent both in the
Lorentzian and the Euclidean regimes, as it actually happens in the
particular case $\Omega^2=1$.

It is a simple exercise to show that, under the conformal transformation
defined by Eq. (16), the lapse and shift functions and the 3-metric of
$ds^2$ and $(ds_P)^2$ are related by
\begin{equation} N=\Omega N_P,  \;\;\;\;N^a=N_P^{\;a},\;\;\;\;q_{ab}=\Omega^2
(q_P)_{ab},\end{equation}
so that the transformation law for the triad is
\begin{equation} e^a_{\;i}=\Omega^{-1}(e_P)^a_{\;i}.\end{equation}
Using Eqs. (18,19) and Eqs. (2-5), it is easy to derive the relations
between the densitized triads and the SO(3) connections of the two
considered four-metrics:
\begin{equation} \tilde{E}^a_{\;i}=\Omega^2 (\tilde{E}_P)^a_{\;i},\;\;\;\;
A_a^{\;i}=(A_P)_a^{\;i}.\end{equation}
Substituting these equations in the gravitational constraints (6-8), and
taking into account the definition (9) of the SO(3) curvature
$F_{ab}^{\;\;\;\;i}$, we obtain that
\begin{equation} {\cal G}_i=\Omega ^2({\cal G}_P)_i=0,\;\; {\cal V}_a=
\Omega^2 ({\cal V}_P)_a=0,\;\;{\cal S}=\Omega^4 {\cal S}_P=0,\end{equation}
with the subindex $P$ denoting evaluation at $(\tilde{E}_P)^a_{\;i}$ and
$(A_P)_a^{\;i}$. Therefore, the gravitational constraints for the Ashtekar
variables $\tilde{E}^a_{\;i}$ and $A_a^{\;i}$ can be equivalently written
by simply evaluating them at the conformally transformed variables.

In the Ashtekar formulation, the classical time evolution is generated by the
Hamiltonian
\begin{equation} H=\int_{\Sigma}d^3x\left(\frac{1}{2}
N_{_{_{\!\!\!\!\!\!\sim}}}\;{\cal S}- i N^a ({\cal V}_a-A_a^{\;i}
{\cal G}_i)+i N^i {\cal G}_i\right)+{\rm Surface}\;{\rm Terms}.\end{equation}
Here, ${\cal G}_i$, ${\cal V}_a$ and ${\cal S}$ are given by Eqs. (6-8),
$N_{_{_{\!\!\!\!\!\!\sim}}}\;=N\,q^{-1/2}$ is the densitized lapse
function, $N^a$ is the shift function and $N^i$ is a SO(3) Lagrange
multiplier [3]. The surface terms in Eq. (22) are introduced to render $H$
finite, and depend on the conditions imposed on the Ashtekar variables
(and the Lagrange multipliers) on the boundary of the 3-manifold $\Sigma$
[2,3]. Since these surface terms do not alter the dynamical equations of
$\tilde{E}^a_{\;i}$ and $A_a^{\;i}$, we will obviate them in the
discussion to follow.

The Hamiltonian (22) leads to the equations of motion [2,3]
\begin{equation} \dot{\tilde{E}\,}\!\,^a_{\;i}=\{\tilde{E}^a_{\;i},H\}=
-i{\cal D}_b\left(N_{_{_{\!\!\!\!\!\!\sim}}}\;\epsilon_i^{\;\,jk}
\tilde{E}^a_{\;j}\tilde{E}^b_{\;k}\right)+{\cal L}_{\bar{N}}
\left(\tilde{E}^a_{\;i}\right)+N^k\tilde{E}^a_{\;j}
\epsilon_{i\;\;\;\;k}^{\;\,j},\end{equation}
\begin{equation}\dot{A}_a^{\;i}=\{A_a^{\;i},H\}=iN_{_{_{\!\!\!\!\!\!\sim}}}\;
\epsilon^{ij}_{\;\;\;\,k}\tilde{E}^b_{\;j}F_{ab}^{\;\;\;\;k}+
{\cal L}_{\bar{N}} A_a^{\;i}+{\cal D}_aN^i,\end{equation}
where the dot denotes time derivative, ${\cal L}_{\bar{N}}$ is the Lie
derivative:
\begin{equation} {\cal L}_{\bar{N}}\left(\tilde{E}^a_{\;i}\right)=
\partial_b\left(N^b\tilde{E}^a_{\;i}\right)-\tilde{E}^b_{\;i}
\partial_bN^a,\;\;\;\; {\cal L}_{\bar{N}}
A_a^{\;i}=N^b\partial_bA_a^{\;i}+A_b^{\;i}\partial_aN^b,\end{equation}
and ${\cal D}_b$ is the derivative operator defined by the Ashtekar
connection:
\begin{equation} {\cal D}_b\left( N_{_{_{\!\!\!\!\!\!\sim}}}\;
\epsilon_i^{\;\,jk}\tilde{E}^a_{\;j}\tilde{E}^b_{\;k}\right)=
\partial_b\left( N_{_{_{\!\!\!\!\!\!\sim}}}\;\epsilon_i^{\;\,jk}
\tilde{E}^a_{\;j}\tilde{E}^b_{\;k}\right)+\epsilon_{il}^{\;\;\,m}A_b
^{\;l}N_{_{_{\!\!\!\!\!\!\sim}}}\;\epsilon_m^{\;\,jk}
\tilde{E}^a_{\;j}\tilde{E}^b_{\;k},\end{equation}
\begin{equation} {\cal D}_a N^i=\partial_a
N^i+\epsilon^i_{\;\,jk}A_a^{\;j}N^k.\end{equation}
Under the conformal transformation (16), the SO(3) Lagrange multiplier $N^i$
remains unchanged, $N^i=N_p^{\;i}$. From the definition of
$N_{_{_{\!\!\!\!\!\!\sim}}}\;$ and Eq. (18), we also conclude that
\begin{equation} N_{_{_{\!\!\!\!\!\!\sim}}}\;=\Omega^{-2}
N_{_{_{\!\!\!\!\!\!\sim}}}\;_P,\;\;\;\;N^a=N_P^{\;a}.\end{equation}
It is then possible to see that the dynamical equations (23,24) are still
valid when evaluated at the Lagrange multipliers and Ashtekar variables
determined  by $(ds_P)^2$.
To prove this statement, it suffices to realize that Eqs.
(1), (20-22) and (28), together with $N^i=N_P^{\;i}$, imply that
\begin{equation} \{(A_P)_a^{\;i}(x),(\tilde{E}_P)^b_{\;j}(y)\}=
\Omega^{-2}i\delta^b_a\delta^i_j\delta^3(x,y),\end{equation}
\begin{equation} H=\Omega^2 H_P.\end{equation}
Thus, the Poisson brackets of $(\tilde{E}_P)^a_{\;i}$ and $(A_P)_a^{\;i}$
with $H$ get a factor $\Omega^2$ from the conformal transformation of the
Hamiltonian, and a factor $\Omega^{-2}$ because of the mo\-dification of
the basic Poisson brackets between the densitized triad and the Ashtekar
connection. Therefore, the equations of motion for $(\tilde{E}_P)^a_{\;i}$
and $(A_P)_a^{\;i}$ coincide with the right-hand side of Eqs. (23,24)
evaluated at the variables associated with the line element $(ds_P)^2$.
This result can also be obtained by substituting Eqs. (20) and (28) (and
$N^i=N_P^{\;i}$) into expressions (23-27).

For $N_P^{\;i}$ and $N_P^{\;a}$ real, the time evolution defined by Eq.
(23) respects the reality of the densitized triad $(\tilde{E}_P)^a_{\;i}$
in the Lorentzian regime ($N_P\in I\!\!\!\,R$) provided that the real part
of $(A_P)_a ^{\;i}$ is equal to the connection
$\Gamma_a^{\;i}(\tilde{E}_P)$ [2,3]. On the other hand, it is
straightforward to check that, for real $N_P^{\;i}$ and $N_P^{\;a}$, the
equations of motion (23,24) are also consistent with the reality of the
densitized triad $(\tilde{E}_P)^a_{\;i}$ and the SO(3) connection
$(A_P)_a^{\;i}$ in the Euclidean regime, i.e., for imaginary lapse
functions $N_P$, for which $i N_{_{_{\!\!\!\!\!\!\sim}}\;P}\in
I\!\!\!\,R$. As we had anticipated, the sections of the complex phase
space of general relativity with real densitized triads up to a constant
complex factor can then be employed to describe, at least in vacuum, both
classical Lorentzian and Euclidean gravity, and the specific regime
considered depends on the values taken by the complexified time
coordinate.

Let us try to find now the sets of reality conditions that correspond to
this family of classical theories. We first note that the reality
conditions for the Lorentzian and Euclidean theories analyzed here should
be essentially the same for each complex conformal factor $\Omega^2$,
because, with $\Omega^2$ fixed, these two types of dynamical regime are
always related by an analytic continuation of the time, and then the
arguments presented in the previous section for the particular case
$\Omega^2=1$ can be applied as well to all these models.

Since the densitized triad $(\tilde{E}_P)^a_{\;i}$ is real in all the
theories that we are studying, we will require
$(\hat{\tilde{E}\;}\!_P)^a_{\;i}$ to be Hermitian as part of the reality
conditions. Thus,
\begin{equation} (\hat{\tilde{E}\;}\!\,^a_{\;i})^{\ast}=\bar{\Omega}^2\left(
(\hat{\tilde{E}\;}\!_P)^a_{\;i}\right)\!^{^{\ast}}=\Omega^{-2}
(\hat{\tilde{E}\;}\!_P)^a_{\;i}=\Omega^{-4}\hat{\tilde{E}\;}\!\,^a_{\;i}.
\end{equation}
The reason to justify this assumption is that it ensures that any possible
real observable constructed only from the metric $(q_P)_{ab}$ will be
represented in the quantum theory by a self-adjoint operator, so that its
expectation values will always be real. The admissible reality conditions
for the SO(3) connection $A_a^{\;i}$ are then severely restricted by the
compatibility of the algebra of commutators (13) with the properties of
the $\ast$-relation (14,15). Defining the triadic extrinsic curvature as
\begin{equation} -i\,\hat{K}_a^{\;i}=\hat{A}_a^{\;i}-\Gamma_a^{\;i}
(\hat{\tilde{E}\;}),\end{equation}
and taking the $\ast$-conjugate  of Eq. (13), we arrive at the result
\begin{equation} \left([i\,\hat{K}_a^{\;i}(x),
\hat{\tilde{E}\;}\!\,^b_{\;j}(y)]\right)^{\ast}=
-\hbar \delta^b_a\delta^i_j\delta^3(x,y)=\Omega^{-4}i
[(\hat{K}_a^{\;i})^{\ast}(x),\hat{\tilde{E}\;}\!\,^b_{\;j}(y)].
\end{equation}
Then, the reality conditions for $\hat{K}_a^{\;i}$ must be of the form
\begin{equation} \left(\hat{K}_a^{\;i}\right)^{\ast}=\Omega^4 \hat{K}_a^{\;i}+
f_a^{\;i}(\hat{\tilde{E}\;}).\end{equation}
Here, $f_a^{\;i}$ is a function of the densitized triad (and its spatial
derivatives) still to be determined.

It is worth remarking that the classical analogue of $\hat{K}_a^{\;i}$
will be, in general, a complex triadic extrinsic curvature, its value
depending on that of the complex lapse function $N_P$. It thus seems
natural not to assume any given reality condition for $\hat{K}_a^{\;i}$,
but to deduce it from the consistency of the algebraic structures.

For the $\ast$-operation to be an involution, we have to require also that
\begin{equation} \hat{K}_a^{\;i}=\left(\hat{K}_a^{\;i}\right)^{\ast\ast}=
\hat{K}_a^{\;i}+\Omega^{-4}f_a^{\;i}(\hat{\tilde{E}\;})+\bar{f}_a^{\;i}
(\Omega^{-4}\hat{\tilde{E}\;}),\end{equation}
and so
\begin{equation} \bar{f}_a^{\;i}(\Omega^{-4}\hat{\tilde{E}\;})=-\Omega^{-4}
f_a^{\;i}(\hat{\tilde{E}\;}).\end{equation}

On the other hand, the reality conditions should guarantee that the
$\ast$-conjugate of the operator constraints do not lead to any new
constraint in the system different from those originally imposed.
Otherwise, the consistency of the quantization procedure will demand the
introduction of additional constraints that were not present in the
classical theory from which one started. This statement has never appeared
in the literature, although it is clear that it has always been implicitly
assumed.

Let us study first the Gauss law (6), which can be equivalently written as
\begin{equation} {\cal G}_i\equiv -i \epsilon_{ij}^{\;\;\;\,k} \hat{K}_a^{\;j}
\hat{\tilde{E}\;}\!\,^a_{\;k}=0.\end{equation}
{}From now on, we will employ in our calculations the symmetric factor
ordering for all the products of the operators $\hat{K}_a^{\;i}$ and
$\hat{\tilde{E}\;}\!\,^a_{\;i}$, even if this ordering is not displayed
explicitly. Applying the $\ast$-operation to Eq. (37), and using reality
conditions (31) and (34), we conclude that, for all possible values of
$\Omega$, ${\cal G}_i^{\;\,\ast}=0$ is equivalent to ${\cal G}_i=0$ if and
only if
\begin{equation}\epsilon_{ij}^{\;\;\;\,k}f_a^{\;j}
(\hat{\tilde{E}\;})\hat{\tilde{E}\;}\!\,^a_{\;k}=0.\end{equation}

Modulo the Gauss law and the Bianchi identities, the vector constraint
(7) can be expressed as [3]
\begin{equation} {\cal V}_a\equiv -iD_b\left(\hat{K}_a^{\;i}
\hat{\tilde{E}\;}\!\,^b_{\;i}-\hat{K}_c^{\;i}
\hat{\tilde{E}\;}\!\,^c_{\;i}\delta^b_a\right)=0,\end{equation}
with $D_b$ the derivative operator compatible with the triad, i.e.,
$D_b(\hat{\tilde{E}\;}\!\,^a_{\;i})=0$ and
\begin{equation} D_b(\hat{K}_a^{\;i})=\partial_b\hat{K}_a^{\;i}+
\epsilon^i_{\,jk}\Gamma_b^{\;j}(\hat{\tilde{E}\;})
\hat{K}_a^{\;k}+\Gamma^c_{\;\;ba}(\hat{\tilde{E}\;})
\hat{K}_c^{\;i}.\end{equation}
{}From Eqs. (31) and (34), it is then easy to check that
${\cal V}_a^{\;\,\ast}=0$ provided that ${\cal V}_a=0$ (and vice versa)
only if $f_a^{\;i}$ satisfies the requirement

\begin{equation}D_b\left( f_a^{\;i}(\hat{\tilde{E}\;})
\hat{\tilde{E}\;}\!\,^b_{\;i}-f_c^{\;i}
(\hat{\tilde{E}\;})\hat{\tilde{E}\;}\!\,^c_{\;i}\delta^b_a \right)=0,
\end{equation}
similar to the constraint (39) on $\hat{K}_a^{\;i}$.

Finally, the scalar constraint (8) can be rewritten, modulo the Gauss
law [3],
\begin{equation} {\cal S}\equiv q^2(\hat{\tilde{E}\;})R(\hat{\tilde{E}\;})+2
\hat{K}_a^{\;i}\hat{K}_b^{\;j}\hat{\tilde{E}\;}\!\,^a_{\;[i}
\hat{\tilde{E}\;}\!\,^b_{\;j]}=0,\end{equation}
where the brackets denote antisymmetrization, $R(\hat{\tilde{E}\;})$ is the
operator associated with the scalar curvature of the 3-metric $q_{ab}$,
and $q^2(\hat{\tilde{E}\;})$ corresponds to the determinant of that metric:
\begin{equation} q^2(\hat{\tilde{E}\;})=\frac{1}{6}\epsilon^{ijk}
\epsilon_{abc}\hat{\tilde{E}\;}\!\, ^a_{\;i}\hat{\tilde{E}\;}\!\,^b_{\;j}
\hat{\tilde{E}\;}\!\,^c_{\;k}.\end{equation}
Taking the $\ast$-conjugate of Eq. (42), and subtracting from it the original
constraint, we arrive after some trivial manipulations at the condition
\begin{equation} (1-\Omega^8) q^2(\hat{\tilde{E}\;})R(\hat{\tilde{E}\;})+
2f_a^{\;i}(\hat{\tilde{E}\;})f_b^{\;j}(\hat{\tilde{E}\;})
\hat{\tilde{E}\;}\!\,^a_{\;[i}\hat{\tilde{E}\;}\!\,^b_{\;j]}=
-4\Omega^4f_a^{\;i}(\hat{\tilde{E}\;})\hat{K}_b^{\;j}
\hat{\tilde{E}\;}\!\,^a_{\;[i}\hat{\tilde{E}\;}\!\,^b_{\;j]},\end{equation}
where we have employed that, under the conformal transformation
$\tilde{E}^a_{\;i}\rightarrow \Omega^{-4}\tilde{E}^a_{\;i}$,
\begin{equation} R(\Omega^{-4}\tilde{E})\equiv R(\Omega^{-4}q_{ab})=
\Omega^{4}R(q_{ab})\equiv\Omega^4R(\tilde{E}).\end{equation}

Since $f_a^{\;i}$ is a function of the densitized triad and its spatial
derivatives, and Eq. (44) includes also the extrinsic curvature
$\hat{K}_a^{\;i}$, such equation can be satisfied without introducing a
new constraint in the system only if it can be derived from the original
gravitational constraints (6-8) for some particular choice of $f_a^{\;i}$.
We note that expression (44) establishes a relation between an homogeneous
function of degree one in the extrinsic curvature and another homogeneous
function of degree equal to zero, both of them dependent on the densitized
triad. Because of this fact, the scalar constraint (42) cannot be used to
eliminate the dependence on the extrinsic curvature in Eq.  (44), for it
can only reduce the order of homogeneity in $\hat{K}_a^{\;i}$ by a
multiple of 2. On the other hand, the vector constraint (39) is
homogeneous of degree one in the extrinsic curvature, and it can only be
employed to substitute some components of the extrinsic curvature as
linear combinations (depending on the triad) of other components and their
spatial derivatives. Therefore, equation (44) is functionally independent
of the scalar and vector constraints.

Finally, the Gauss law (37) implies that $\hat{\tilde{K}\;}\!\,^i_{\;j}=
\hat{K}_a^{\;i}\hat{\tilde{E}\;}\!\,^a_{\;j}$ is symmetric in the SO(3)
indices. Rewritten then the right-hand side of Eq. (44) as
\begin{equation} -2\Omega^4\left(\tilde{f}^i_{\;i}
(\hat{\tilde{E}\;})\hat{\tilde{K}\;}\!\,^j_{\;j}
-\tilde{f}^i_{\;j}(\hat{\tilde{E}\;})\hat{\tilde{K}\;}\!\,^j_{\;i}\right),
\end{equation}
and taking into account that $\tilde{f}^i_{\;j}(\hat{\tilde{E}\;})=
f_a^{\;i}(\hat{\tilde{E}\;})\hat{\tilde{E}\;}\!\,^a_{\;j}$ is symmetric
because of condition (38), we conclude that the sole way in which relation
(44) can be satisfied, for $\hat{\tilde{K}\;}\!\,^j_{\;i}$ a generic SO(3)
symmetric tensor operator that is functionally independent of the
densitized triad, is that both sides of the considered equation vanish. In
particular, expression (46) must be identically zero, whichever the values
of the symmetric extrinsic curvature may be. In order to verify this
requirement, the function $\tilde{f}^i_{\;j}$ must vanish, as one can
check after some trivial calculations. Then,\linebreak
$f_a^{\;i}(\hat{\tilde{E}\;})=\tilde{f}^i_{\;j}(\hat{\tilde{E}\;})
\hat{E_{_{_{\!\!\!\!\!\!\sim}}\;}}^j_{\;a}$ has to be equal to zero (at
least for non-degenerate metrics; if we want to extend our conclusions to
the degenerate case, we have to assume the continuity of $f_a^{\;i}$ as a
function of the densitized triad).

Obviously, for $f_a^{\;i}=0$, the additional conditions (36), (38) and
(41) are immediately fulfilled. The only consistency demand that we have not
discussed yet is the vanishing of the left-hand side of Eq. (44) at
$f_a^{\;i}=0$:
\begin{equation}  (1-\Omega^8) q^2(\hat{\tilde{E}\;})R(\hat{\tilde{E}\;})=0.
\end{equation}
This condition can be satisfied, for instance, by restricting our
attention to flat minisuperspace models. For these models, we have proved
then that (at least in vacuum) it is possible to adopt sets of reality
conditions of the form
$(\hat{\tilde{E}\;}\!\,^a_{\;i})^{\ast}=\Omega^{-4}
\hat{\tilde{E}\;}\!\,^a_{\;i}$
and $(\hat{A}_a^{\;i})^{\ast}=-\Omega^4A_a^{\;i}$ with $\Omega$ any
complex constant ($|\Omega|=1$). In the general case, however, we will
have that $ q^2(\tilde{E}) R(\tilde{E})$ is different from zero. Thus, in
the full theory of gravity one is forced to demand that $\Omega^8=1$. In
other words, once one assumes that the densitized triad
$(\hat{\tilde{E}\;}_P)^a_{\;i}$ (that is real in the classical theory) is
Hermitian, the compatibility of the reality conditions with the algebraic
structures and constraints of general relativity restricts the complex
conformal factor to be such that $\Omega^8=1$. For those theories with
other constant values of $\Omega$, the densitized triad
$\hat{\tilde{E}\;}\!\,^a_{\;i}$ cannot be a Hermitian operator up to a
complex constant factor, and any classical real observable that might be
constructed from the 3-metric will have in general genuine complex
expectation values in the corresponding quantum theory.

Let us focus then our discussion on the case $\Omega^8=1$. Since reality
conditions (31) and (34), with $f_a^{\;i}=0$, depend just on the fourth
power of $\Omega$, it will suffice to consider two possible values for the
conformal factor: $\Omega^2=1$ and $\Omega^2=-i$. The case $\Omega^2=1$
was analyzed in Sec. 2; it corresponds to the Lorentzian sector of general
relativity, with reality conditions given by
$(\hat{\tilde{E}\;}\!\,^a_{\;i})^{\ast}=\hat{\tilde{E}\;}\!\,^a_{\;i}$ and
$(\hat{A}_a^{\;i})^{\ast}=-\hat{A}_a^{\;i}+
2\Gamma_a^{\;i}(\hat{\tilde{E}\;})$.
On the other hand, we will argue in the next section that the reality
conditions for $\Omega^2=-i$ can be associated in a natural way with the
Euclidean sector of pure gravity.

\section{Reality Conditions for Euclidean Gravity}

When the conformal factor $\Omega^2$ is purely imaginary, reality conditions
(31) and (34) can be rewritten
\begin{equation} (\hat{\tilde{E}\;}\!\,^a_{\;i})^{\ast}=
-\hat{\tilde{E}\;}\!\,^a_{\;i}
,\;\;\;\;(\hat{A}_a^{\;i})^{\ast}=\hat{A}_a^{\;i},\end{equation}
where we have used the definition (32) for the triadic extrinsic
curvature. The second of these relations affirms that the Ashtekar
connection is Hermitian, as one would expect it to happen in Euclidean
gravity. In fact, it is not difficult to show that reality conditions (48)
are simply the direct translations of the complex conjugation relations
for real Euclidean gravity in the Ashtekar formalism.

To prove this statement, let us study the classical theory obtained from
Lorentzian general relativity by a Wick rotation of the time and a constant
and purely imaginary conformal transformation. In the Ashtekar formulation,
the gravitational action can be expressed [3]
\begin{equation} S=\int dt\left[\int _{\Sigma}d^3x(-i\tilde{E}^a_{\;i}
\dot{A}_a^{\;i})-H'\right] ,\end{equation}
where $H'$ coincides formally with the Hamiltonian $H$ given by Eq. (22)
without surface terms. This action is equal on shell to the
Hilbert-Einstein action of general relativity [3,11], which is real in the
Lorentzian regime. Under a Wick rotation of the time, the Lorentzian
action transforms into $i$ times the action for Euclidean gravity
[12,15,18], the latter being real for Euclidean 4-metrics. On the other
hand, the transformation laws (20) and (30) (still valid for $H'$) imply
that, for the theory with conformal factor $\Omega^2=-i$, the Ashtekar
action is exactly $-i$ times the action of the non-transformed variables.
Therefore, in the classical theory with imaginary conformal factor and
Wick rotated time, the gravitational action (49) turns out to coincide on
shell with the real classical action of pure Euclidean gravity. In terms
of the Ashtekar variables, the Euclidean theory is described by the
densitized triad $(\tilde{E}_P)^a_{\;i}=i\,\tilde{E}^a_{\;i}$ and the
connection $(A_P)_a^{\;i} =A_a^{\;i}$, which, in agreement with our
previous remarks, are both real in the classical theory and Hermitian when
considered as operators. Thus, reality conditions (48) ensure that all
real observables constructed from the Euclidean 3-metric and extrinsic
curvature are self-adjoint in the quantum theory.

Another topic that we want to address is whether reality conditions (10)
and (48), that correspond, respectively, to Lorentzian and Euclidean
gravity, determine at the end of the day unitarily equivalent quantum
theories. If the answer were in the positive, one could use the simple
reality conditions (48) to quantize general relativity, and obtain from
that quantum theory physical predictions for Lorentzian gravity [20].
However, we will prove that, under very general assumptions, this is not
indeed the case.

Reality conditions (10) and (48) will lead to equivalent quantum theories
if there exists an isomorphism between their respective $\ast$-algebras of
elementary operators that leaves invariant the gravitational constraints.
In the following, we will use the symbol $I$ to denote this isomorphism,
and the subindex 0 to refer to those operators that belong to the
Euclidean $\ast$-algebra.

The isomorphism $I$ must satisfy the requirements
\begin{equation} I(\hat{X}_0\hat{Y}_0)=I(\hat{X}_0)I(\hat{Y}_0),\end{equation}
\begin{equation} I(\hat{X}_0^{\;\ast})=(I(\hat{X}_0))^{\;\ast},\end{equation}
\begin{equation} I([\hat{X}_0,\hat{Y}_0])=[I(\hat{X}_0),I(\hat{Y}_0)],
\end{equation}
with $\hat{X}_0$ and $\hat{Y}_0$ any two operators. In particular, from the
first of these equations we get that $I(\hat{1}_0)=\hat{1}$.

Since the operators $i(\hat{\tilde{E}\;}_0)^a_{\;i}$ and
$\hat{\tilde{E}\;}\!\,^a_{\;i}$ are both Hermitian, and both of them
describe classical real densitized triads, it seems natural to assume that
\begin{equation} I((\hat{\tilde{E}\;}_0)^a_{\;i})=-i
\hat{\tilde{E}\;}\!\,^a_{\;i}.\end{equation}
The consistency condition $I(\hat{\tilde{E}\;}_0\!\,^{\;\ast})=
(I(\hat{\tilde{E}\;}_0))^{\ast}$ is then automatically verified.

The isomorphism for the connection operator $(\hat{A}_0)_a^{\;i}$ can be
derived from Eqs. (13) and (52,53), for they imply that
\begin{equation} I\left([(\hat{A}_0)_a^{\;i}(x),
(\hat{\tilde{E}\;}_0)^b_{\;j}(y)]
\right)=[I((\hat{A}_0)_a^{\;i}(x)), -i\,\hat{\tilde{E}\;}\!\,^b_{\;j}(y)]=
-\hbar \delta^b_a\delta^i_j\delta^3(x,y),\end{equation}
and thus we must have
\begin{equation} I((\hat{A}_0)_a^{\;i})=i\,\hat{A}_a^{\;i}+
g_a^{\;i}(\hat{\tilde{E}\;}),\end{equation}
with $g_a^{\;i}$ a yet undetermined function of the densitized triad.
Recalling the definition of the triadic extrinsic curvature, and employing
Eq. (50) and
$\Gamma_a^{\;i}(-i\,\hat{\tilde{E}\;})=\Gamma_a^{\;i}(\hat{\tilde{E}\;})$,
we obtain that
\begin{equation}I((\hat{K}_0)_a^{\;i})=i\,\hat{K}_a^{\;i}+f_a^{\;i}
(\hat{\tilde{E}\;}),\end{equation}
where $f_a^{\;i}\equiv i\, g_a^{\;i}-(1+i)\Gamma_a^{\;i}$. It can be seen
then that $I(\hat{K}_0^{\;\ast})=(I(\hat{K}_0))^{\ast}$ provided that
$\bar{f}_a^{\;i}(\hat{\tilde{E}\;})=-f_a^{\;i}(\hat{\tilde{E}\;})$.

Equations (53) and (56) fix the isomorphism between $\ast$-algebras up to
a function $f_a^{\;i}$. As we have commented above, this isomorphism must
also leave invariant the constraints of general relativity, that is, Eqs.
(37), (39) and (42). Given the similarity of Eqs. (53) and (56) with the
$\ast$-relations (31) and (34), it is not difficult to realize that one
can parallel the discussion of the previous section about the equivalence
of the gravitational constraints and their $\ast$-conjugate by simply
regarding the isomorphism $I$ as a $\ast$-operation of the type (31) and
(34) with $\Omega^4=i$, and conclude therefore that, for the isomorphism
to respect the constraints, one must have $f_a^{\;i}=0$ and
\begin{equation} q^2(\hat{\tilde{E}\;})R(\hat{\tilde{E}\;})=0,\end{equation}
this last equation arising from condition (48) evaluated at $\Omega^4=i$.
So, except in some particular cases, i.e., in flat minisuperspace models,
there exists no isomorphism between the $\ast$-algebras associated with
reality conditions (10) and (48) such that it maps the Euclidean Hermitian
triad $i(\hat{\tilde{E}\;} _0)^a_{\;i}$ into the Lorentzian triad
$\hat{\tilde{E}\;}\!\,^a_{\;i}$ and preserves in addition the
gravitational constraints. Then, the sets of reality conditions that
correspond to Euclidean and Lorentzian gravity determine inequivalent
quantum theories, and it seems impossible to regain Lorentzian physics
from the quantum theory obtained by using the Euclidean reality
conditions.

\section {Conclusions}

We have shown that the set of reality conditions that are usually
associated with Euclidean gravity, i.e., that the densitized triad and the
Ashtekar connection be Hermitian, are inconsistent with the algebra of
commutators in the Ashtekar formulation of general relativity. We have
also argued that the reality conditions are invariant under a Wick
rotation of the time, and thus the quantum theories that describe
Lorentzian gravity and its Euclideanized version are in fact completely
equivalent.

We have considered then a family of classical theories that are related to
Lorentz\-ian and Euclidean gravity in vacuum by a constant conformal
transformation. If one requires that the classical real triad is
represented by a Hermitian operator in these theories, the compatibility
of the $\ast$-operation with the algebra of commutators and constraints
restricts the admissible complex conformal factors to be either real or
purely imaginary. For real conformal factors, one recovers the Lorentzian
section of general relativity and its well-known set of reality
conditions. In the case of a purely imaginary conformal factor, the
classical theory can be identified as real Euclidean gravity, and the
corresponding set of reality conditions guarantee that any real observable
derived from the Euclidean 3-metric and extrinsic curvature is
self-adjoint in the quantum theory. Therefore, these reality conditions
can be interpreted as those associated with Euclidean general relativity.
Explicitly, the Euclidean Ashtekar connection must be Hermitian, and the
densitized triad has to be represented by an anti-Hermitian operator.

Finally, we have proved that the Lorentzian and the Euclidean sets of
reality conditions lead to inequivalent quantum theories once one
identifies the operators that describe the classical 3-metric in these two
different quantizations. It thus seems impossible to extract Lorentzian
physical predictions from the quantum theory determined by the Euclidean
reality conditions, which would have been much simpler to impose when
proceeding to the quantization of general relativity.

\vspace*{.5cm}
{\bf Acknowledgements}

\vspace*{.4cm}
The author is greatly thankful to A. Ashtekar, F. Barbero
and D. Marolf for helpful discussions. He wants to
thank also the Center of Gravitational Physics and Geometry at the
Pennsylvania State University for warm hospitality. This work was
supported by funds provided by the Spanish Ministry of Education and
Science Grant No. EX92-06996911.

\end{document}